\documentclass[prl,twocolumn,nofootinbib]{revtex4-1}
\usepackage{graphicx}
\usepackage{color}
\usepackage{dcolumn}
\usepackage{bm}
\usepackage{amssymb}
\usepackage[bottom]{footmisc}
\usepackage{footnote}
\usepackage{lipsum}
\usepackage{wrapfig}
\usepackage{sidecap}
\renewcommand{\thefootnote}{\fnsymbol{footnote}}

\newcommand\blfootnote[1]{%
\begingroup
\renewcommand\thefootnote{}\footnote{#1}%
\addtocounter{footnote}{-1}%
\endgroup}
\begin{document}
\title{Electric-field control of magnetism in few-layered van der Waals magnet}
\author{Zhi Wang,$^{1,2*}$ Tong-Yao Zhang,$^{3,4*}$  Mei Ding,$^{5*}$ Baojuan, Dong,$^{1,2*}$ Yan-Xu Li,$^{3,4}$ Mao-Lin Chen,$^{1,2}$ Xiao-Xi Li,$^{1,2}$ Yong Li,$^{1,2}$ Da Li,$^{1,2}$ Chuan-Kun Jia,$^{5}$ Li-Dong Sun,$^{6}$  Huaihong Guo,$^{7}$ Dong-Ming Sun,$^{1,2}$ Yuan-Sen Chen,$^{3,4\dagger}$ Teng Yang,$^{1,2\dagger}$ Jing Zhang,$^{3,4}$ Shimpei Ono,$^{8}$ Zheng Vitto Han,$^{1,2\dagger}$ and Zhi-Dong Zhang$^{1,2}$}

\affiliation{$^{1}$Shenyang National Laboratory for Materials Science, Institute of Metal Research, Chinese Academy of Sciences, Shenyang 110016, China}
\affiliation{$^{2}$School of Material Science and Engineering, University of Science and Technology of China, Anhui 230026, China}
\affiliation{$^{3}$State Key Laboratory of Quantum Optics and Quantum Optics Devices, Institute of Opto-Electronics, Shanxi University, Taiyuan 030006, P. R. China}
\affiliation{$^{4}$Collaborative Innovation Center of Extreme Optics, Shanxi University, Taiyuan 030006, P.R.China}
\affiliation{$^{5}$College of Materials Science and Engineering, Changsha University of Science $\&$ Technology, Changsha, 410114, China}
\affiliation{$^{6}$State Key Laboratory of Mechanical Transmission, School of Materials Science and Engineering, Chongqing University, Chongqing, China}
\affiliation{$^{7}$College of Sciences, Liaoning Shihua University, Fushun, 113001, China}
\affiliation{$^{8}$Central Research Institute of Electric Power Industry (CRIEPI), Materials Science Research Laboratory, 2-6-1 Nagasaka, Yokosuka, Kanagawa, Japan}


\maketitle
\blfootnote{\textup{*} These authors contribute equally.}

\blfootnote{$^\dagger$Corresponding to: yuansen.chen@googlemail.com, yangteng@imr.ac.cn, and vitto.han@gmail.com}

\textbf{Manipulating quantum state via electrostatic gating has been intriguing for many model systems in nanoelectronics. When it comes to the question of controlling the electron spins, more specifically, the magnetism of a system, tuning with electric field has been proven to be elusive. Recently, magnetic layered semiconductors have attracted much attention due to their emerging new physical phenomena. However, challenges still remain in the demonstration of a gate controllable magnetism based on them. Here, we show that, via ionic gating, strong field effect can be observed in few-layered semiconducting Cr$_{2}$Ge$_{2}$Te$_{6}$ devices. At different gate doping, micro-area Kerr measurements in the studied devices demonstrate tunable magnetization loops below the Curie temperature, which is tentatively attributed to the moment re-balance in the spin-polarized band structure. Our findings of electric-field controlled magnetism in van der Waals magnets pave the way for potential applications in new generation magnetic memory storage, sensors, and spintronics.}

 \section{Introduction}

The family of two-dimensional (2D) van der Waals (vdW) materials hold great promise for both fundamental physics and future applications.\cite{Geim_Roadmap, Neto_Science} One of the major characteristics of vdW materials is that, by diminishing the dimensionality into a 2D limit, Coulomb screening reduces significantly, which allows the constructions of new concept nano-electronic transistors or sensors via gate-tuning the Fermi levels of the resulted intrinsic semi-metals or semiconductors.\cite{Cory_NatNano,SpinValve_PRL,Duan_review,Iwasa_Science}.

Among the variety of vdW materials, layered magnetic compounds are of particular interest thanks to the enriched spin-related physics in the system.\cite{Sachs_PRB_2013, McGuire_CM_2015, Tian_2DM_2016, Sivadas_PRB_2015, McGuire_Crystals_2017, Lee_APLMater_2016, May_PRB_2016, Liu_SciRep_2016, Kurumaji_PRL_2011, PRL_Prediction, NatNano_Prediction} For example, the long-standing Mermin-Wagner restriction \cite{MW_Theorem} has been recently lifted in the monolayer limit in CrI$_{3}$, which shows Ising ferromagnetism despite of the zero out-of-plane dimensionality\cite{XD_Xu_Nature, Xiang_Nature}. Previous studies on few-layered vdW magnets have also manifested much promising physical phenomena, including the observation of spin-resolved Raman modes,\cite{FePS_Raman_NanoLett, FePS_Raman_2DMat} the spin waves measured by neutron scattering,\cite{SpinWave_CrSiTe} potential applications in spinvlaves,\cite{APL_SpinValve} as well as the synthesis of new magnetic semiconductors by doping non-magnetic vdW crystals.\cite{ZhongmingWei_NC, Fluorided_BN} Recent advances in the peculiar tunnelling magneto-resistance in CrI$_{3}$ further reveals that atomically thin vdW magnetic materials may serve as potential magnetic sensors.\cite{Morpurgo_CrI3, Pablo_CrI3, XiaodongXu_CrI3}

   \begin{figure*}[ht]
   \includegraphics[width=0.8\linewidth]{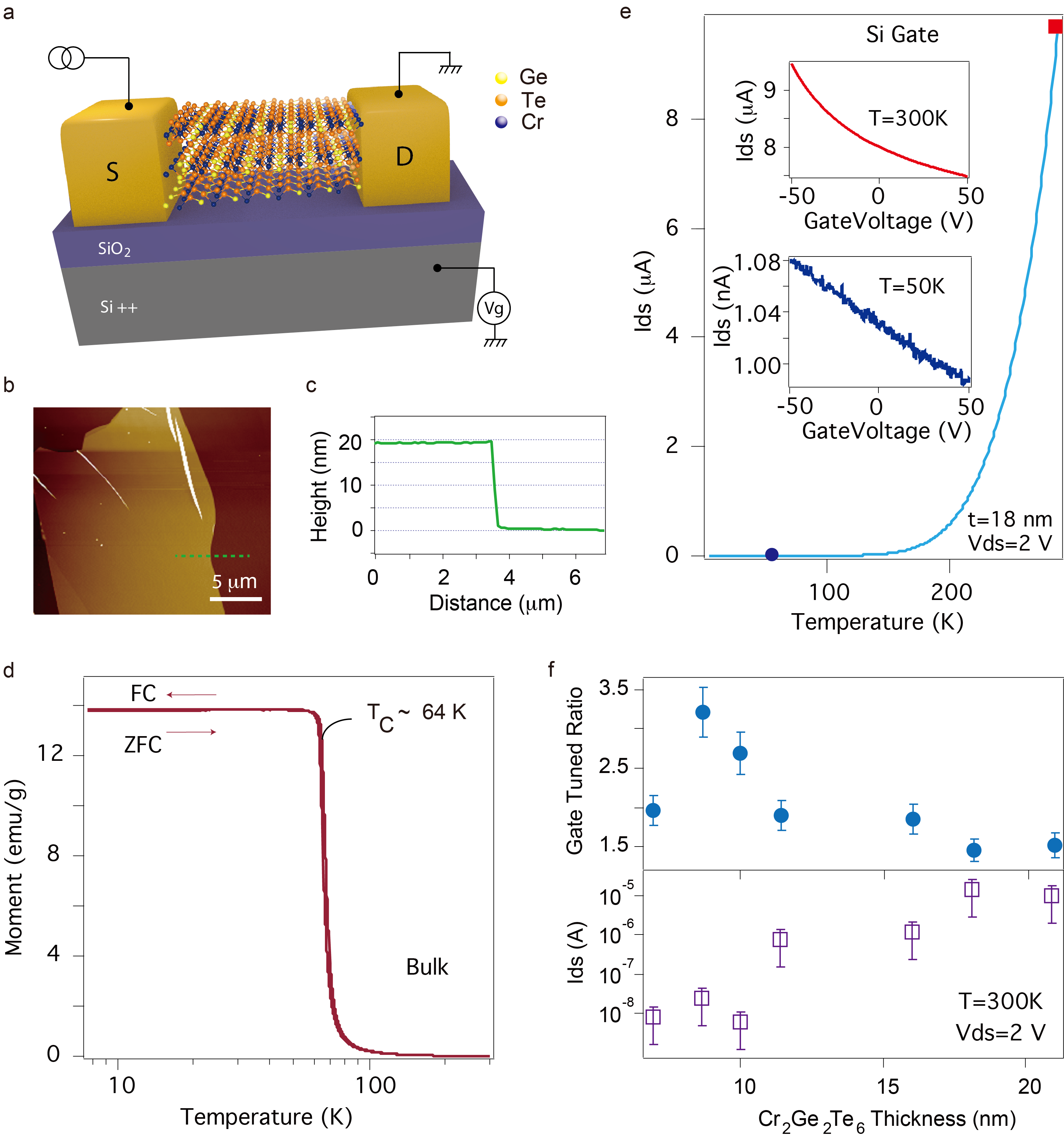}
   \caption{\textbf{Characterization of Cr$_{2}$Ge$_{2}$Te$_{6}$ and its FETs with Si gate.} (a) Schematics of a Au-contacted few-layered Cr$_{2}$Ge$_{2}$Te$_{6}$ device. (b) AFM image of a typical Cr$_{2}$Ge$_{2}$Te$_{6}$ flake of about 19 nm in thickness, with its height profile along the green dashed line plotted in (c). (d) Magnetization of bulk Cr$_{2}$Ge$_{2}$Te$_{6}$ as a function of temperature with zero-field cooling and field cooling processes. (e) Ids as a function of temperature for a typical 19 nm Cr$_{2}$Ge$_{2}$Te$_{6}$ device, with Vds=2V and V$_{g}$=-50V. Insets: upper and lower panels show field effect curves at 300 K and 50 K, respectively. Red and blue colors of the inset curves correspond to the solid square and circle marks indicating each position in the Ids-T curve. (f) A statics on Cr$_{2}$Ge$_{2}$Te$_{6}$ FET devices with different layer thicknesses at room temperature. Upper panel shows the gate tuned ratio calculated from the ratio of Ids at V$_{g}$=-50 V and +50 V; lower panel shows the Ids at V$_{g}$=0 V. All measurements were done with Vds=2V.
   }
   \end{figure*}

However, albeit most of the vdW magnets behave as semiconductors, very few reports have been conducted on field effect transistors (FETs) based on them.\cite{CrSiTe_JMCC, HanWei_2DMat} Especially, studies that utilize the electric field as a knob to effectively tune their magnetism is thus far missing. It is known that electric-field control of magnetization is key for future applications in spintronics, e.g., spin transistors.\cite{Spin_transistor, HideoOhno_Review} To date, to reach this goal, multiple approaches have been developed, including multiferroic heterostructures, \cite{Cherifi_NM2014} thin metals,\cite{Chiba_NM2011, Weisheit_Science_2007} multilayered magnetic thin films,\cite{WeigangWang_NM2012, Maruyama_NN2009, SongCheng_AM2015} and diluted magnetic semiconductors.\cite{Hideo_Nature_DMS2000, Chiba_Science2003}

In this work, we show that few-layered  semiconducting Cr$_{2}$Ge$_{2}$Te$_{6}$ devices can remain conducting and gate tunable below their ferromagnetic Curie temperature. Micro-area Kerr measurements at low temperatures were then carried out on the basis of those Cr$_{2}$Ge$_{2}$Te$_{6}$ transistors with ionic liquid gating, which enabled tunable magnetization loops at different gate doping. The observed behavior of gate-tuned magnetism in few-layered Cr$_{2}$Ge$_{2}$Te$_{6}$ paves the way for future spintronic applications using vdW magnet as a platform.

  \begin{figure}
  \includegraphics[width=1\linewidth]{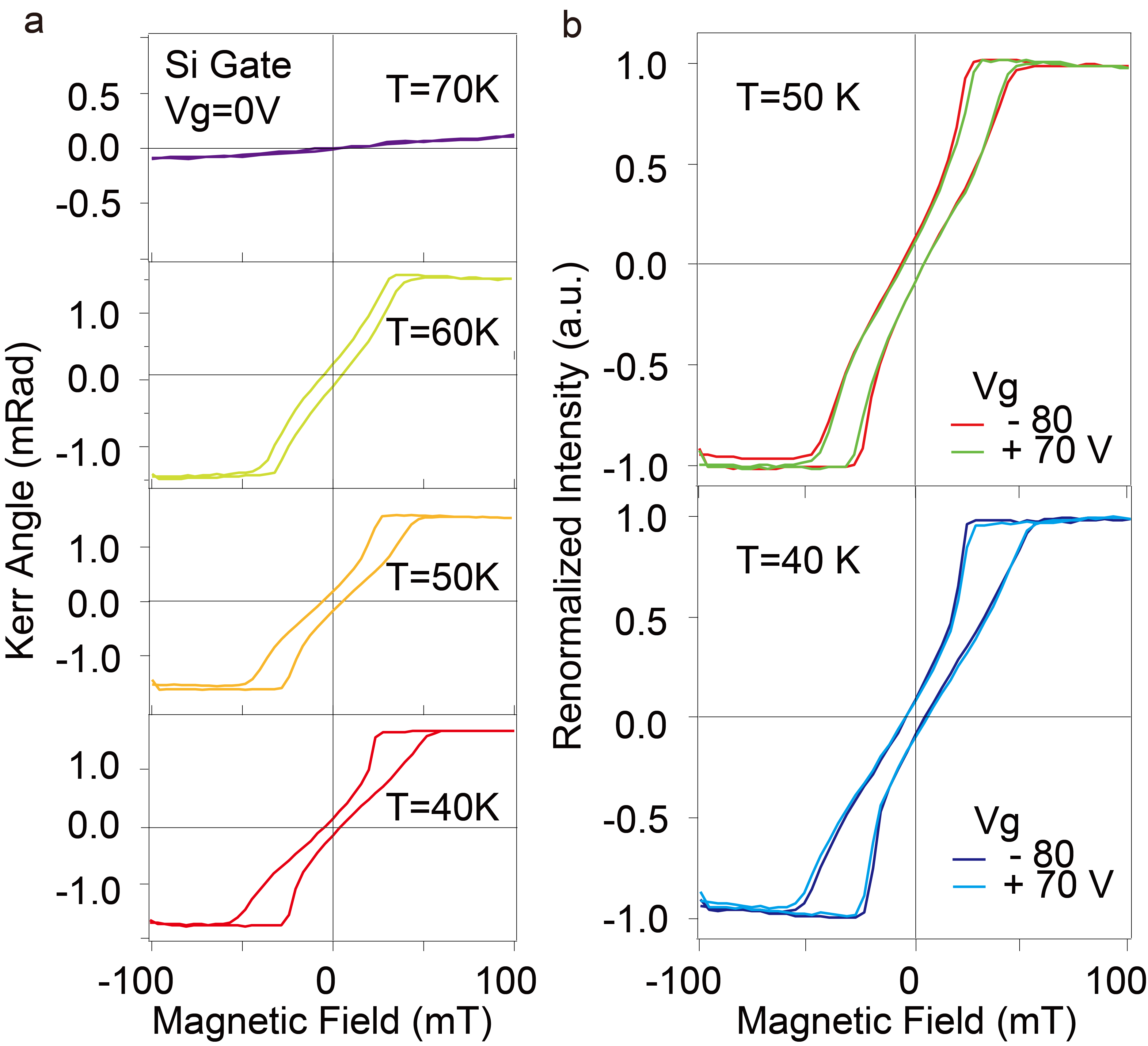}
  \caption{\textbf{Micro-area Kerr measurements on Cr$_{2}$Ge$_{2}$Te$_{6}$ FETs with Si Gate.} (a) Magnetization loop measured in a micro-area of 1 $\mu$m diameter at different temperatures. (b) Comparison (T=50 K and 40 K) of magnetization loops at $V_{g}=-80 V$ and +70 V, respectively. No clear doping effect is seen in Si gated devices within the largest voltage window applied in the gate leakage limit of less than 1 nA.}
  \label{fig:fig2}
  \end{figure}

\begin{figure*}
\includegraphics[width=0.8\linewidth]{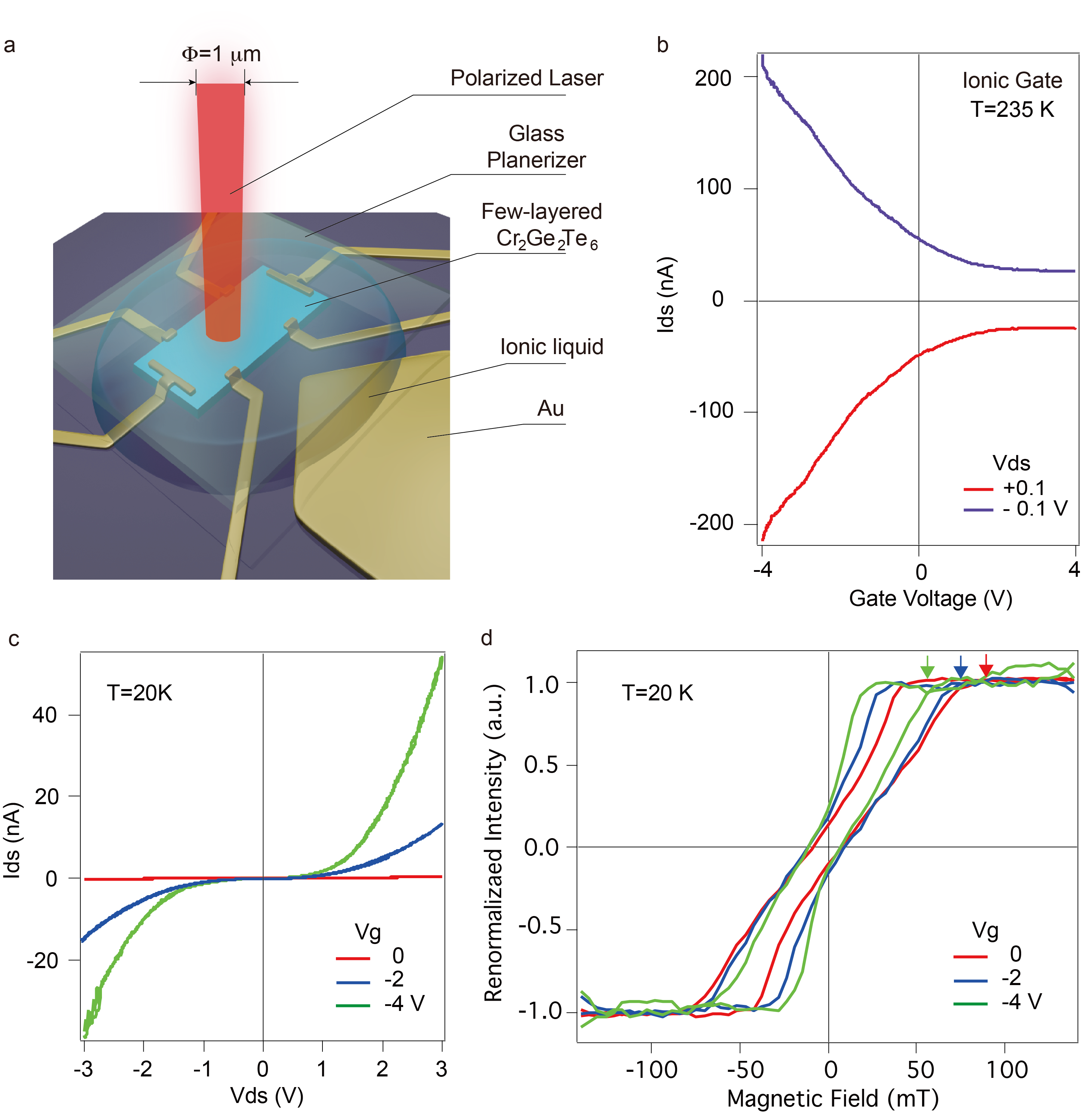}
\caption{\textbf{Electrical transport and magnet porperties Cr$_{2}$Ge$_{2}$Te$_{6}$ FETs with ionic Gate.} (a) Schematic image of the experimental setup for Kerr measurement of the micron-sized device using ionic liquid. (b) Field effect curves at 235 K with $V_{ds}=$ -0.1 and +0.1 V, respectively. (c) IV characteristics of the same device with fixed ionic gate voltages of 0, -2 and -4V, that is measured at 20 K, respectively. (d) Renormalized Kerr angle measured at 20 K and with fixed ionic gate voltages of 0, -2 and -4V, respectively. Colored arrows indicate the H$_{s}$ for the loops measured at each V$_{g}$.  Notice that the ionic liquid freezes below 220 K, therefore the curves are taken for several different cooling-downs.}

\label{fig:fig3}
\end{figure*}

\begin{figure*}
\includegraphics[width=0.8\linewidth]{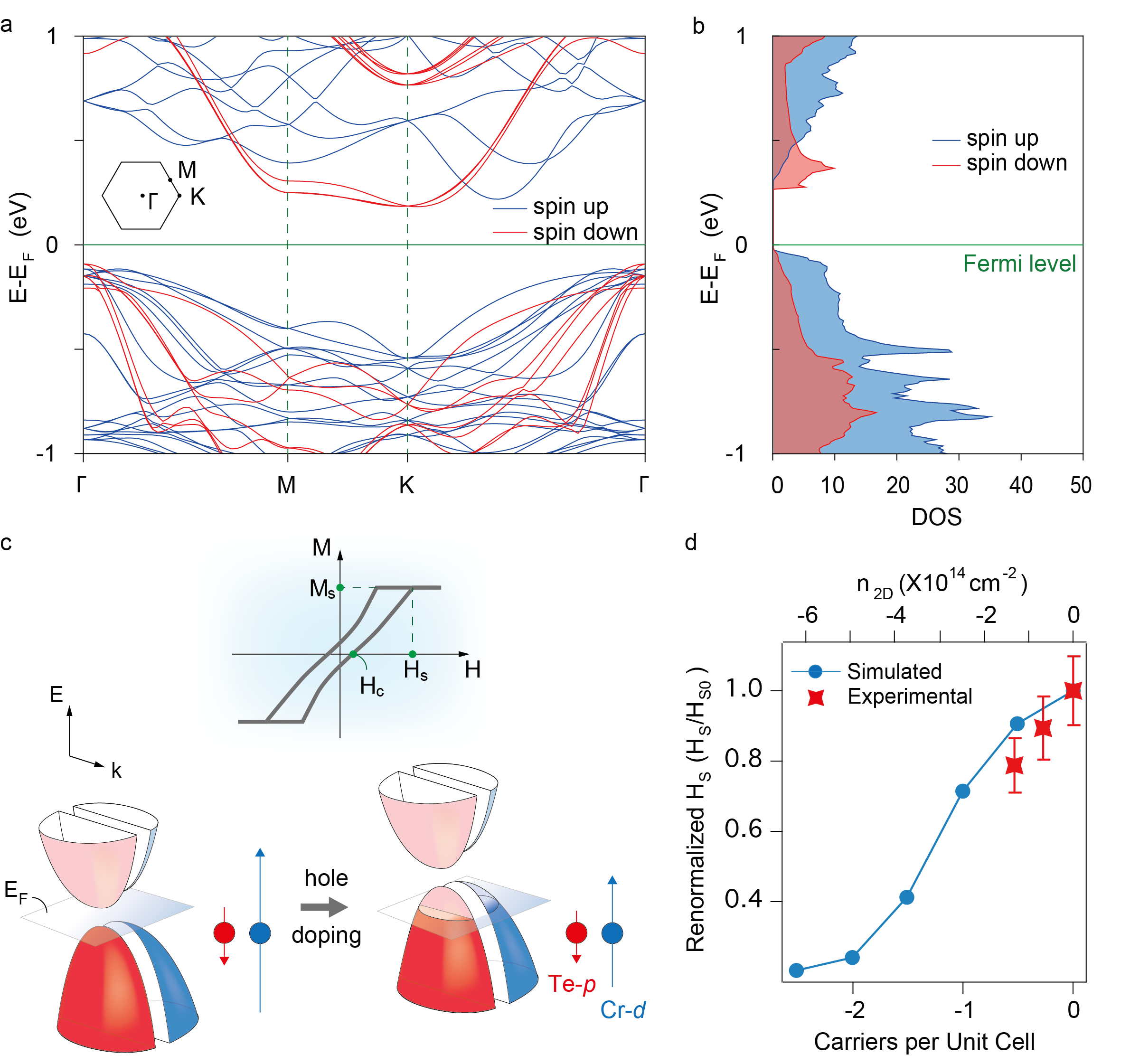}
\caption{\textbf{First principles simulations.}  (a) Spin polarized band structure of few-layered Cr$_{2}$Ge$_{2}$Te$_{6}$. (b) Calculated electron density of states. (c) Upper panel: Schematic magnetization reversal loop. Lower panel: Schematic picture of the spin re-alignment model via Fermi level shifting during electrostatic doping. Red and blue represent for the down and up spins from Te-$p$ orbitals, and Cr-$d$ orbitals, respectively. Semi-transparent and solid colors are empty and occupied bands, respectively. (d) Blue solid circles with solid line is the simulated evolution of net spin as a function of carrier density. Red solid star symbols are experimental data in this work}

\label{fig:fig4}
\end{figure*}

\section{Results}
\textbf{Fabrication of few-layered Cr$_{2}$Ge$_{2}$Te$_{6}$ FETs.} Single crystal Cr$_{2}$Ge$_{2}$Te$_{6}$ was prepared via the Te self-flux method and was confirmed via x-ray diffraction (see Methods and Fig.S1 in the Suppl. Info.). We then applied the scotch tape method to exfoliate the bulk and deposited few-layered Cr$_{2}$Ge$_{2}$Te$_{6}$ onto 285 nm thick silicon oxide grown on heavily doped silicon wafers for further FET fabrications (Fig. 1a). Atomic Force Microscope (AFM) scan of such typical flakes is shown in Figure 1 (b)–(c). Before measuring few-layered Cr$_{2}$Ge$_{2}$Te$_{6}$, we performed zero-field-cooled (ZFC) and field-cooled (FC) thermal magnetization curves of its bulk. As shown in Fig.1d, the Curie temperature is determined to be around 64 K with a 1000 Gauss magnetic field applied perpendicular to the $ab$ plane of the bulk crystal (Fig.S2).

First, we characterize the Cr$_{2}$Ge$_{2}$Te$_{6}$ FETs with Cr/Au (5nm/50nm) electrodes. For a device with thickness $t\sim$ 18 nm, with the source-drain voltage $V_{ds}$ = 2V, one obtains source-drain current $I_{ds}$ at the order of a few $\mu$A. $I_{ds}$ as a function of temperature is recorded in Fig. 1e with a constant $V_{ds}$ of 2V. It is seen that $I_{ds}$ drastically decrease with lowering the temperature, which follows a $T^{-1/4}$ law (Fig. S3). Moreover, IV characteristics of the devices show an Ohmic to Semiconducting transition with lowering the temperature, as shown in Fig. S4. These behaviours are of a typical semiconductor, agree with previously reported \cite{HanWei_2DMat, Morpurgo_CrI3}. Field effect curves of the devices show typical p-type FET behavior. As shown in the inset of Fig. 1e, the gate tuned ratio (defined as the ratio of $I_{ds}$ between gate voltage $V_{g}$=-50 V and +50 V) decrease from about 1.5 to 1.08 at 300 K and 50 K, respectively, indicating a weakened gate efficiency at low temperatures.

It is noticed that, even at room temperature, few-layered thin vdW magnetic materials often already exhibit faint current (at the order of nA) with a few volts applied between source and drain electrodes.\cite{Morpurgo_CrI3, HanWei_2DMat, CrSiTe_JMCC} It therefore indicates that a rather insulating state will be expected at low temperature, giving rise to difficulties for transport measurements below the Curie temperatures. For example, we found that the Cr$_{2}$Ge$_{2}$Te$_{6}$ device with 8 nm thickness turned into insulator below 150 K (Fig.S5), which is yet much higher than its Curie Temperature (T$_{C}$).

To retain the gate tunability of Cr$_{2}$Ge$_{2}$Te$_{6}$ transistors at low temperatures, we investigated a statistics on a series of devices to meet the trade off between the layer thickness and electrical conductivity. Figure S6-S8 shows the optical image of the measured devices and their field effect curves at room temperature. It is seen that the devices exhibit linear IV characteristics at all gate voltages at room temperature. However, as summarized in Fig. 1f, clear increase of resistance (drop of $I_{ds}$) is seen with decreasing the layer thickness. At a fixed $V_{ds}=2V$, The device with 20 nm thickness exhibited $I_{ds} \sim$ 10 $\mu$A, which is 4 orders of magnitudes higher than that of those with 6-10 nm thickness. Meanwhile, along with the augmented $I_{ds}$, an reduced gate tuned ratio is observed in thicker devices (1.5 for 20 nm device as compared to 3.5 for 8 nm device). It is noteworthy that Cr$_{2}$Ge$_{2}$Te$_{6}$ flakes thinner than 6 nm start to show poorer gate tunability, which may be caused by the air instability as often seen in other 2D crystals.\cite{WLLiu_2DM} As a result, in order to obtain conducting and gate-tunable few-layered Cr$_{2}$Ge$_{2}$Te$_{6}$ below T$_{C}$, it is favourable to study the devices with thickness $t > 15 nm$. In the following context, we will focus on typical devices with $t$= 18-20 nm.

\bigskip

\textbf{Kerr characterization of Cr$_{2}$Ge$_{2}$Te$_{6}$ FETs with Si gate.} Magneto-optic Kerr effect is known as an method to probe the magnet moment according to the changes to light reflected from a magnetized surface. It has been widely used for surface science, including the recently thrived 2D vdW magnetic materials.\cite{Morpurgo_CrI3, HanWei_2DMat, XD_Xu_Nature, Xiang_Nature} Here, we use an ultra-high sensitivity Kerr setup with a low temperature vacuum cryostat to investigate the magnetization of few-layered Cr$_{2}$Ge$_{2}$Te$_{6}$ transsitors, while monitoring their electrical transport $in-situ$. The optical diagram and the measuring protocols are described in the Suppl. Info. By using an objective close to sample surface, spot size of about 1 $\mu$m diameter is achieved in the incident laser with a wavelength of 800 nm.   Figure 2a shows a micro-area Kerr measurements at different temperatures of a Cr$_{2}$Ge$_{2}$Te$_{6}$ device with thickness of about 19 nm. It is seen that at T=70 K, the sample behaves as paramagnetic, in good agreement with the ZFC-FC characterization of it bulk form. When decreasing temperature below 60 K, it shows ferromagnetic loops, with enhanced saturation field at lower temperature (Fig. S10).

To investigate the possible solid gate tuning of the Kerr signal, we carried out field effect measurement of the studied device, as shown in Fig. S11. A maximum gate range was determined from -80 V to +70 V, to the limit which keeps gate leakage negligible. According to the model of capacitive coupling of a Si gate with a SiO$_{2}$ dielectric layer (285 nm used here), the charge carrier concentration induced per unit area, per unit Volt, is about 7.6 $\times 10 ^{10} $cm$ ^{-2}$V$^{-1}$.\cite{Yuanbo_QHE2005} Therefore, the total charge doping that took effect was at the order of 10$^{13} $cm$^{-2}$. However, even with the largest possible gate voltages applied in the Cr$_{2}$Ge$_{2}$Te$_{6}$ transistors, no obvious tuning of Kerr signal was observed at T$< $T$_{C}$, as shown in Fig. 2b.

\bigskip

\textbf{Kerr characterization of Cr$_{2}$Ge$_{2}$Te$_{6}$ FETs with ionic gate.} In the following, we discuss the measurement of Cr$_{2}$Ge$_{2}$Te$_{6}$ FETs using an ionic liquid as gate dielectric. Ionic liquid such as are often used in gate tuning semiconductor conducting channels, because of the molecular in the liquid can form an electric double-layer (EDL) that significantly reduces the thickness as compared to conventional solid dielectric materials.\cite{Iwasa_NM2010, Hongtao_JACS2010, Morpurgo_NanoLett2015} It is therefore also referred to as EDL transistors.

Figure 3a shows a schematic image of a 20 nm thick Cr$_{2}$Ge$_{2}$Te$_{6}$ flake on a Si/SiO2 substrate contacted by a Cr/Au (5/50 nm) electrodes. Together with the contacts, a large area pad acting as gate was defined close to the sample. To form the ionic-gated FET, a small droplet of ionic liquid DEME-TFSI was placed onto the device and the gate pad, followed by a glass cap that covers the whole area. Only this way it will then allow the incident of laser for further Kerr measurement.  Before starting cooling, the device was left in vacuum at room temperature at a pressure of $\sim 10 ^{-4}$ mbar for a few hours. It was then cooled down to 235 K (above the freezing point of DEME-TFSI \cite{Iwasa_NanoLett2012}) first to test the field effect curves.

As shown in Fig. 3b, within a chemical window of $\pm$4 V (see more details in the Suppl. Info.), the sample is stable and a strong field effect can be obtained, with a gate tuned ratio of about 10 (defined as the ratio of $I_{ds}$ between gate voltage $V_{g}$=-4 V and +4 V). IV characteristics show linear behaviors at all gate voltages at 235 K, as shown in Fig. S13. Notice that the field effect curve is stabilized after several thermal cycles (Suppl. Info.) as there is a surface potential reconfiguration process until a stable state is reached.\cite{Morpurgo_NanoLett2015} Once the field effect curve is stabilized, we started cool down the sample below T$_{C}$ for Kerr measurements. Figure 3c shows IV characteristics of the same device with fixed ionic gate voltages of 0, -2 and -4V, that is measured at 20 K, respectively. A semiconducting output curve is seen at low temperature, but with a much higher $I_{ds}$ compared with those obtained in the Si gated devices for similar Cr$_{2}$Ge$_{2}$Te$_{6}$ flake thickness.

Strikingly, unlike the Si gated Cr$_{2}$Ge$_{2}$Te$_{6}$ FETs (Fig. 2b), renormalized magnetization loop of the devices gated by ionic liquid can be largely tuned as shown in Fig. 3d. Take the -4V$_{g}$ loop for example, the saturation field H$_{s}$, indicated by gray and black arrows in Fig. 3d, can be modified by a factor of 2 compared to the values of the loop measured at V$_{g}$=0V.  For the ionic liquid DEME-TFSI used here, the gate induced carrier density $n_{2D}$ has been widely studied before, which reaches about 1.0 and 0.75 $\times 10 ^{14}$cm$^{-2}$ for electrons and holes at $\left | V_{g} \right |$  = 3 V, respectively, which are usually about 100 times higher per volt than that found in solid-state FETs with 285 nm oxides.\cite{Morpurgo_NanoLett2015, Iwasa_NanoLett2012}


\section{Discussion}


In the following, we compare the measured data with first-principles simulations. Given that our experimental Cr$_{2}$Ge$_{2}$Te$_{6}$ flakes are more than 18 nm in their thicknesses, we here invoke a 3-layer bulk phase of Cr$_{2}$Ge$_{2}$Te$_{6}$ for computational simplicity. The lattice parameters are drawn from the Rietveld refinement of our X-ray diffraction data and are consistent with the Springer Materials database.\cite{SMdatabase} In Fig.~\ref{fig:fig4}a,b, we show electron band structure and density of states (DOS). Cr$_{2}$Ge$_{2}$Te$_{6}$ is a ferromagnetic semiconductor, and its spin majority bands are mainly contributed by $d$ orbital of Cr atoms and spin minority bands from the $p$ orbital of Te atoms in both valence and conduction bands near the band gap, as seen from projected DOS in Fig.S15. Hole doping (electron doping) by electrostatic gating will shift the Fermi level into valence (conduction) band by depleting (filling) more Cr-$d$ orbitals than Te-$p$ orbitals, which leads to a reduced net magnetic moment and saturated moment under magnetic field, as indicated in the schematic picture in Fig.~\ref{fig:fig4}c.

In order to see how coercivity field H$_c$ and saturation field H$_s$ changes with electrostatic doping, we calculate the dependence of energy on both net spin moment and doping level by combining the density function theory and the fixed spin moment method\cite{Moruzzi86}. As shown in Fig.S16, for all studied hole doping level from 0.0 to 1.5 carriers/unit cell, energy profile exhibits a bi-stable electronic ground state at m$_{gs}$ $\neq$ 0, exhibiting a typical ferromagnetism. As mentioned earlier, m$_{gs}$ decreases with increasing the doping level. Meanwhile, both coercivity field (H$_c$) and saturation field (H$_s$) are extracted from the calculated energy profile, namely H$_{c,s}$ = $\frac{\partial E}{\partial m} \mid_{c,s}$ (definitions of H$_{s}$ and H$_{c}$ are indicated in the upper panel of Fig.4c). More details will be given in the methods part.

Weisheit et al. reported that magnetocrystalline anisotropy K$_{U}$ of FePt and FePd 2 nm thin films can be reversibly modified by an applied electric field with the EDL technique.\cite{Weisheit_Science_2007} In their case, the measurable that changes with ionic gate doping is mainly the coercivity $H_{c}$ that differs for around 3 $\%$ (compared to 50 $\%$ $\delta$H$_{s}$ obtained in this work). They assumed the $\delta$H$_{c}$ to be directly proportional to the K$_{U}$, and can be evaluated with respect to K$_{U}$ energies derived from electronic structure calculations.\cite{Weisheit_Ref9} Here, the $H_{c}$ obtained in the few-layered Cr$_{2}$Ge$_{2}$Te$_{6}$ devices are very weak and the trend of modification is below the detection sensitivity of our set-up. We therefore will mainly focus on the electric-field modification of H$_{s}$, as well as the amplitude of Kerr angle ($i.e.$, the saturated magnetization, M$_{s}$) recorded at different gate doping.

In Fig.~\ref{fig:fig4}d, we show the carrier density dependence of the saturation field H$_s$ (normalized by the zero-doping H$_s$(0)) in blue solid line. H$_s$ goes down along with the increasing hole carrier concentration, which agrees well with the experimental data shown in red star. In contrary, the coercivity field H$_c$ has a very weak doping level dependence, as shown in Fig. S17. Meanwhile, shown in Fig. S18, experimentally a reduced M$_{s}$ is observed with increasing hole doping, which is in qualitative agreement with the simulated result in Fig. S17. The doping-induced effect can be comprehended as follows. From the schematic picture in Fig.~\ref{fig:fig4}c, saturation field H$_s$ needed for reversing the minority spin is expected to go down with increased hole concentration. While the coercivity field H$_s$ is strongly relevant to magneto-crystalline anisotropy energy (MAE). Our calculation shows that MAE between in-plane and out-of-plane magnetization direction is quite small (at least one order of magnitude smaller than soft iron.) and MAE has no obvious dependence on doping level of interest here.



\bigskip

In conclusion, we have demonstrated that few-layered semiconducting Cr$_{2}$Ge$_{2}$Te$_{6}$ devices can serve as transistors that exhibit a gate-tuned modification of magnetism, which is obtained from the micro-area Kerr measurements below the ferromagnetic Curie temperature with ionic liquid gating. The observed behavior of gate-tuned magnetism in few-layered Cr$_{2}$Ge$_{2}$Te$_{6}$ may be attributed to a re-balance of spin polarized band structure while tuning its Fermi level. Our findings therefore prove that vdW magnet can be a promising platform that may open  further opportunities for future applications in spin-transistors.

\section{Methods}

Single crystal Cr$_{2}$Ge$_{2}$Te$_{6}$ was prepared via the Te self-flux method. Raw material powders with stoichiometric ratio of Cr (purity 99.9$\%$): Ge (purity 99.9$\%$):Te (purity 99.99$\%$)=1:4:20 were mixed and kept at 950 $^{o}$C for 6 h. The mixture was then cooled at the rate of 2 $^{o}$C h$^{-1}$, followed by a centrifuge at 500 $^{o}$C.

As depicted in the Suppl. Info., Kerr rotation was performed to monitor the electrically tunable out-of-plane magnetization of the sample investigated. A cw Ti-Sapphire laser of 100 kHz bandwidth was used to generate linearly-polarized light and the probe energy was fixed at 1.550 eV for all measurements. By means of a lens with a numerical aperture of 0.50, the Gaussian beam was tightly focused with a sigma width of 1.5 $\mu$m on the sample surface. A balanced photo diode bridge is adopted to sensitively monitor the polarization change of the probe beam due to the polar magneto-optic Kerr effect (MOKE). The sample was mounted strain-free in Faraday geometry in a helium-free cryostat, $i.e.$, both the external magnetic field applied $B_\textrm{\scriptsize ext}$ and the light propagation vector are along the out-of-plane direction of the sample. The sample can be moved by an x-y-z piezo stage with an accuracy of 200 nm, and the sample temperature can be set from 4.0 K to 350 K. To characterize the coercive field and saturation magnetization of the sample, MOKE loop measurements were performed continuously by scanning longitudinal magnetic field from a negative $\left | B_\textrm{\scriptsize ext} \right |$ to a positive $\left | B_\textrm{\scriptsize ext} \right |$, and backwards to -$\left | B_\textrm{\scriptsize ext} \right |$, which produced the MOKE loops recorded at various experiment conditions.

The simulations in this work are carried out by using the first-principles density functional theory as implemented in the \textsc{VASP} code\cite{VASP}. Projector augmented wave (PAW) pseudopotentials~\cite{PAWPseudo} and the Perdew-Burke-Ernzerhof (PBE)~\cite{PBE} functional are respectively used to describe electron-ion interaction and electronic exchange-correlation interaction. The Brillouin zone of the primitive unit cell is sampled by $6{\times}6{\times}2$~$k$-points~\cite{Monkhorst-Pack76}. We adopt $500$~eV as the electronic kinetic energy cutoff for the plane-wave basis and $10^{-6}$~eV as
the criterion for reaching self-consistency. Fixed-spin moment method \cite{Moruzzi86} is used to simulate energy-moment behavior. Saturation and coercivity fields are obtained by H$_{c,s}$ = $\frac{\partial E}{\partial m} \mid_{c,s}$. Coercivity point are chosen at the inflex m point (maximum of $\mid \frac{\partial E}{\partial m} \mid$) between m = 0 and m$_{gs}$ and saturation point chosen at the m point where 20-p valence electrons giving rise to saturated magnetization m$_s$, with p as the doping level. The magneto-crystalline anisotropy energy is calculated by using a more dense k mesh, i.e., $12{\times}12{\times}4$~$k$-points.

\section{\label{sec:level1}ACKNOWLEDGEMENT}
This work is supported by the National Natural Science Foundation of China (NSFC) with Grant 11504385 and 51627801, and is supported by the National Key R$\&$D Program of China (2017YFA0206302). Z.D. Zhang acknowledges supports from the NSFC with grant 51331006 and the CAS under the project KJZD-EW-M05-3. T.Yang acknowledges supports from the Major Program of Aerospace Advanced Manufacturing Technology Research Foundation NSFC and CASC, China (No. U1537204). The work in Shanxi University is financially supported by the NSFC (Grant No. 61574087) and the Fund for Shanxi ``1331 Project” Key Subjects Construction (1331KSC).

\section{Author contributions}
Z.H. and Z.-D.Z. conceived the experiment and supervised the overall project. Z.W. fabricated the samples. Z.W., T.-Y.Z., Y.-X.L., Y.-S.C., and Z.H. carried out experimental measurements; S.O. provided the ionic liquid and advised on the experiment; B.-J.D. and T.Y. conducted the theoretical simulations. The manuscript was written by Z.H., Y.-S.C., and T. Y. with discussion and inputs from all authors.

\bibliographystyle{naturemag}

\end{document}